\documentclass{kasper}
\usepackage{xspace}
\usepackage[bottom]{footmisc}

\makeatletter
\@ifundefined{url}{\newcommand{\url}[1]{$#1$}}{}
\makeatother

\newcommand{\halfbps}{\hbox{$1/2$-BPS}\xspace}
\newcommand{\quarterbps}{\hbox{$1/4$-BPS}\xspace}
\DeclareMathOperator{\tr}{tr}

\begin{document}
\preprintnumber{CERN-TH/2001-186\\DAMTP-2001-64\\hep-th/0107164}
\date{July 19th, 2001}
\email{k.peeters, m.zamaklar@damtp.cam.ac.uk}
\title{Motion on moduli spaces with potentials}
\author{Kasper Peeters$^{1}$ and Marija Zamaklar$^{2}$}
\address{1}{CERN\\
TH-division\\
1211 Geneva 23\\
Switzerland}
\address{2}{DAMTP/CMS\\
Cambridge University\\
Wilberforce road\\
Cambridge CB3 0WA\\
United Kingdom}
\maketitle
\begin{abstract}
  In the limit of small velocities, the dynamics of half-BPS
  Yang-Mills-Higgs solitons can be described by the geodesic
  approximation. Recently, it has been shown that quarter-BPS states
  require the addition of a potential term to this approximation. We
  explain the logic behind this modification for a larger class of
  models and then analyse in detail the dynamics of two
  five-dimensional dyonic instantons, using both analytical and
  numerical techniques. Nonzero-modes are shown to play a crucial
  role in the analysis of these systems, and we explain how these
  modes lead to qualitatively new types of dynamics.
\end{abstract}
\maketoc
\begin{sectionunit}
\title{Introduction}
\maketitle

Much about the kinematics of supersymmetric solitons is known by
virtue of the fact that their properties are related to the
superalgebra of the underlying field theory. On the other hand the
dynamical aspects of these objects are much harder to analyse.
Extracting information about the dynamics of solitons by direct
analysis of exact, time dependent solutions to the full equations of
motion is generically too complicated.  Instead, one uses the
approximate but very powerful \emph{moduli space technique} first
introduced by \dcite{Manton:1982mp} for the study of the dynamics of
four-dimensional Yang-Mills-Higgs magnetic monopoles.

Manton's method is based on the idea that the dynamics of solitons can
be studied, to lowest order in the velocity, by a quasi-static
approach. The set of static solutions, modulo small gauge
transformations, is parametrised by a number of collective coordinates
or moduli, and the space they span is called \emph{moduli space}.  At
each moment of time the system is assumed to be infinitesimally close
to a particular static configuration on this space. An effective
action which tells us how the system evolves from one static
configuration to another can be constructed as follows. 

One first promotes the parameters in the static solution to be slowly
varying functions of time. In order to deal only with the physical
degrees of freedom, one has to ensure that this configuration, with
time-dependent parameters, also satisfies the Gauss' constraint at the
lowest order in the velocity and at each point in time. Very often
this requires adding compensating gauge transformation terms to the
static solution. This time-dependent configuration does by
construction not change the potential energy, and the deviation from
the original static solution is therefore called a
\emph{zero-mode}. If the initial (static) configuration was minimising
the potential energy, then the zero-modes are deformations which will
make the slowly moving system evolve through the set of states with
the same, minimal value of the potential energy.

By inserting now this time-dependent solution into the action,
expanding it to lowest order in the velocity and integrating over the
space-like coordinates, one arrives at an effective action for the
moduli of the static configuration. This action is just an ordinary
non-linear world-line sigma model with the moduli space as the target
space. The metric on it is naturally induced from the field theory
action\footnote{\label{f:AHconstruction} Remarkably, the metric on the
  most studied example of two SU(2) monopoles was actually never
  obtained this way.  Instead, \dcite{Atiyah:1985dv} were able to fix
  the metric completely using only symmetry arguments. This method has
  clear calculational advantages, as the zero-modes for this case are
  very complicated. In general situations, however, there are not
  enough symmetries and one has to resort to the zero-mode method
  described above.}, and the dynamical evolution of the system is
described by geodesics on the moduli space. Although strictly speaking
this approach is valid only in the limit of very small velocities, an
analysis of the dynamics of two Yang-Mills-Higgs monopoles by
\dcite{Manton:1988bn} shows that this method is surprisingly accurate.
Radiation effects contribute only a tiny correction at speeds up to a
considerable fraction of the speed of light.

An interesting feature that the moduli space method has brought to
attention is the fact that four-dimensional \halfbps \emph{dyonic}
solitons can be interpreted as zero-mode excitations of an underlying
purely magnetically charged soliton.  In particular the Julia-Zee
dyon~\cite{juli1} arises naturally in this approach as an excited
't\,Hooft-Polyakov monopole. This can be seen by expanding the
Julia-Zee field theory solution to lowest order in the electric
charge. One is then left with an \emph{unmodified} 't\,Hooft-Polyakov
monopole and a non-vanishing $A_{0}$ component of the gauge field.
Since the electric part of the energy of the dyon is a purely
``kinetic'' contribution to the total energy, the question arises as
to what it is that is moving in these solutions. It turns out that the
$A_{0}$ component of the solution is precisely equal to the global
gauge rotation zero-mode of the starting 't\,Hooft-Polyakov monopole,
with the velocity equal to the electric charge.  The direction of this
additional zero-mode is compact, which accounts for the quantisation
of the electric charge.  One concludes that the moduli space
approximation for monopoles \emph{automatically} incorporates dyons,
and dynamical processes can lead to charge exchange.
\medskip
 
Recent developments in the study of dualities in gauge theory and
string theory have uncovered novel four and five-dimensional
\quarterbps dyonic solutions, starting with the work of
\dcite{Bergman:1997yw}. In contrast to the \halfbps dyons discussed
above, the dynamics of these solitons is naturally described on the
moduli space of \emph{approximately static configurations}.  This is
strongly tied to the way in which those solutions are constructed in
the first place.  In the cases known so far, the \quarterbps dyons can
be seen as deformations of \halfbps electrically uncharged solutions,
obtained by turning on an additional Higgs field. The equations one
has to solve are an \emph{unmodified} BPS equation for the underlying
\halfbps configuration, together with an equation for the additional
scalar field in the background of the underlying \halfbps soliton.
The latter has a unique solution, and thus provides a one-to-one map
between the moduli spaces of the two types of solitons. However,
turning on the additional Higgs makes the system unstable, because the
Higgs field introduces a new attractive force which is unbalanced. By
further adding an additional electric charge, a compensating force is
introduced which prevents the configuration from collapsing. The
resulting solution is dyonic in the sense that it carries both the
charge of the original \halfbps soliton as well as the electric charge
introduced to prevent collapse.

The instability of the intermediate step, where the Higgs is turned on
but the electric charge is still zero, can alternatively be seen
through the appearance of a \emph{potential on moduli space}. This
potential can in principle be derived just like the metric part of the
action, but most of what is known is actually based on
supersymmetry. Since rather general arguments show that the potential
has to be written as the square of a Killing vector on moduli space,
the analysis of \dcite{Alvarez-Gaume:1983ab} can be used to argue that
the orbits of the Killing vector have to preserve the three complex
structures in order for it to be compatible with the four
supersymmetries. This observation, due to~\dcite{Tong:1999mg} and
elaborated on in many other papers, has led to a determination of
explicit potentials for several dyon systems in SU(3)
Yang-Mills-Higgs. The general structure of the effective action, and
its supersymmetric extension, has been derived in a series of
papers by~\ddcite{Bak:1999vd}{Bak:1999da} and
\ddcite{Gauntlett:1999vc}{Gauntlett:2000ks}.

The electric charge again arises from a dynamical effect on the moduli
space, just as in the case of the Julia-Zee dyon. However, the charge
deformation is now a zero-mode of an \emph{approximately static}
configuration.  Even though the starting configuration is unstable,
the system becomes stable once the electric charge zero-mode is
added. In fact, this excited configuration corresponds to the
\emph{exact} \quarterbps solution. Heuristically, the stabilisation
arises because the centrifugal force due to the motion of the system
in the compact direction is balanced against the force produced by the
potential. In summary, these new static \quarterbps dyons are
identified with closed geodesics of constant energy on the moduli
space of \halfbps solitons with a potential\footnote{For lack of a
better alternative we keep referring to the solutions to the equations
of motion as ``geodesics''. Of course these solutions are not
necessarily paths of extremal length when the effect of the potential
is included.}.  The conserved quantity specifying different geodesics
is the angular momentum, and its value is identified with the electric
charge carried by the dyon.
\medskip

Although the origin and the form of the potential are thus rather well
understood, it is important to realise that inclusion of a potential
leads one \emph{beyond} the strict moduli space approximation. The
difference becomes most apparent when one explicitly tries to analyse
the \emph{dynamics} of such models, and the present paper is an
attempt to do so. The addition of the potential clearly only makes
sense when its magnitude and slope are small compared to other scales
in the problem. In this regime, the unstable configurations solve the
equations of motion to lowest order in the Higgs expectation value,
and small deformations of this approximate solution can be studied.
The picture one should have in mind is very similar to the picture of
the \halfbps case: the system is restricted to move at the bottom of a
potential valley, which is surrounded by steep mountains. The
\emph{key difference} is that in the \quarterbps case the valley is
slightly tilted. The smallness of the slope is required to prevent the
system from gaining enough kinetic energy to start climbing the
mountains.

With such a tilted valley, the deformations one has to consider are
not just the zero-modes (i.e.~the modes which keep the potential
energy constant) but also the destabilising modes (i.e.~the modes
which change the potential by moving slowly down the tilted valley).
In general it is very difficult to include nonzero-modes in the
dynamics, because there is no simple way to determine which of the
nonzero-modes are most relevant. While a systematic approach has been
described by~\dcite{Manton:1988ba}, there are not very many models for
which this has been worked out in detail (see \dcite{Manton:1996ex}
for one application and further references). In the present case,
however, one can argue that an appropriate set of nonzero-modes
consists of the lifted zero-modes of the underlying $1/2$-BPS soliton.

Given this insight, one can study the explicit solutions to the
equations of motion of the effective action, and the symmetries that
govern them. We will show how the inclusion of nonzero-modes sometimes
leads to the appearance of several separated scales in the low-energy
dynamics, a new feature which does not occur for the \halfbps models.
Instead of ignoring the nonzero-modes, one can actually integrate them
out explicitly, and the resulting dynamics is qualitatively different
from the dynamics predicted by just the zero-modes.  Throughout the
paper we will use the explicit example of two dyonic instantons, which
are solitons in 4+1~dimensional Yang-Mills-Higgs theory. They have the
advantage over other $1/4$-BPS solitons that the metric on moduli
space is not too complicated, and can be constructed explicitly from a
paper by~\dcite{Osborn:1981yf}.  \medskip

The plan of the paper is then as follows. First, we recall the details
of \quarterbps (multi) dyonic solitons in 4+1~dimensional
Yang-Mills-Higgs theory. In section~\ref{s:modes} we then exhibit the
appearance of a potential on moduli space, illustrated by this
particular example. The dynamics of a single dyonic instanton can be
studied analytically, which we do in
section~\ref{s:singledynamics}. The main part of our paper, contained
in section~\ref{s:doubledynamics}, is concerned with the study of the
two-soliton sector. We first derive the metric and potential, and then
perform a detailed analysis of the dynamics. Qualitatively new
features are found in section~\ref{s:nume} and the final section is
concerned with the separation of scales in the low-energy theory. We
end with some comments on work in progress concerning \quarterbps
dyons in 3+1~dimensions.

Two appendices have been added. One discusses technical details
of the two-instanton moduli space, while the other one summarises our
conventions on quaternions and lists a number of useful expressions
involving them.
\end{sectionunit}

\begin{sectionunit}
\title{Dyonic instantons and moduli spaces}
\maketitle
\begin{sectionunit}
\title{The static dyonic instanton solution}
\maketitle
\label{s:staticDI}
After the general introduction we will, in the rest of the paper,
specialise to a concrete example, namely the study of dyonic
instantons.  Before we discuss their dynamics, let us first introduce
the static solution.  The model under consideration is the five
dimensional Yang-Mills-Higgs action, given by
\begin{equation}
\label{e:action}
S = -\int\!{\rm d}^5 x\, \Big(\tfrac{1}{4} \tr\left(F_{\mu\nu}F^{\mu\nu}\right) + 
\tfrac{1}{2} \tr\left( D_\mu\Phi D^\mu \Phi\right)\Big) \, ,
\end{equation}
with $\mu=(0,(i=1..4))$.  This model can be extended to a
supersymmetric one, but we will not need the details of this extension
in the present paper.

For static solutions the equations of motion reduce to
\begin{align}
\label{e:eoms}
D_j F^{ji} + [A_0, D^i A_0] - [\Phi,D^i \Phi] &= 0\,, \\[1ex]
\label{e:gauss}
D_j F^{j0} - [\Phi,D^0 \Phi] &= 0 \,, \\[1ex]
\label{e:laplace}
D_i D^i \Phi &= 0\, .
\end{align}
The first line clearly shows that by choosing $A_0=\pm\Phi$ the last
two terms cancel and the remainder of the equation is solved by an
(anti)~self-dual field-strength~\footnote{For Julia-Zee dyons all three terms
mix and the relation between $A_0$ and $\Phi$ is less rigid.}. One can
alternatively see this by rewriting the Hamiltonian in Bogomol'nyi
form, which leads to the following BPS~equations,
\begin{align}
\label{e:BPS1}
F_{ij} &= s\tfrac{1}{2}\epsilon_{ijkl} F^{kl}\, , \\[1ex]
\label{e:BPS2}
F_{0i} &= s'D_{i}  \Phi \, ,
\end{align}
with $s$ and $s'$ two arbitrary signs (corresponding to the sign of
the instanton and electric charge respectively).  Combining the second
BPS equation with the Gauss law and going to the $A_{0}= \pm\Phi$ gauge
one arrives again at the equation of motion~\eqn{e:laplace} for the
Higgs field, which is just a Laplace equation in the background of the
instanton.  Hence as advertised, the equations for the \quarterbps
dyon reduce to two equations: a BPS equation for the instanton and a
Laplace equation for the scalar. An important observation is that for
each instanton solution to~\eqn{e:BPS1} there is a unique solution
to~\eqn{e:laplace}. Hence there is \emph{a one to one map between the
  moduli space of instantons and the moduli space of dyonic
  instantons.}

For instantons with identical group embedding angles, the solution to
these equations is given by \dcite{Lambert:1999ua}, extended by
\dcite{Eyras:2000dg} and given a brane interpretation by one of the
authors in~\cite{Zamaklar:2000tc}. For $s=-1$ one obtains,
\begin{equation}
\label{e:sol-single}
\begin{aligned}
A_0     &= s'\frac{v}{H}\, T^3\, ,  \\[1ex]
A_i     &= \eta_{ij}^{a}\,\partial_j(\ln H)\, T^a\, , \\[1ex]
\Phi    &= -\frac{v}{H}\, T^3 \, ,
\end{aligned}
\end{equation}
where the function $H$ is given by
\begin{equation}
H       = 1+ \sum_{i} \frac{\rho_{i}^2}{|x^i-y^i|^2}\, .
\end{equation}
The symbols $\eta^a_{ij}$ are the 't~Hooft symbols and the matrices
$T^a$ are Pauli matrices.  This solution is characterised by two
conserved charges: the instanton number
\begin{equation}
I = \frac{1}{2} \int\!{\rm d}^4x\, \epsilon^{ijkl} \tr(F_{ij} F_{kl}) = N\, ,
\end{equation}
and the electric charge
\begin{equation}
q =  s'\frac{1}{v} \int\!{\rm d}^4x\, \tr(D_i \Phi)^2 = 4 \pi^2 v s'
\sum_{i} \rho_{i}^2 \ .
\end{equation}
This solution can be extended to an arbitrary ADHM configuration for
the four-dimensional subspace, as the solution to the Higgs field in
the instanton background has been derived in terms of ADHM data by
\dcite{dore1}. We will make use of this later when we study the
two-dyon dynamics.
\end{sectionunit}

\begin{sectionunit}
\title{Modes, zero-modes and the moduli space approximation}
\maketitle
\label{s:modes}

We now want to analyse the dynamics of the dyonic instanton solutions.
The solution which was described in the previous section is an
\emph{exact, static} solution of the equations of motion. In principle
one could try to construct the zero-modes of this configuration and
the corresponding effective action for them. This method was followed
by~\dcite{Bak:1999sv} for \quarterbps monopoles. Our approach is more
in spirit of Manton's original treatment~\cite{Manton:1985hs} of
Julia-Zee dyons\footnote{It is an interesting open problem to
construct a moduli space effective action for two Julia-Zee dyons
\emph{starting} from the dyonic solution, and understand to what
extent such an action agrees with the one constructed with monopoles
as starting point. This would be the analogue of starting from two
dyonic instantons in the model under consideration here.}.  Namely, we
will try to make a link between dyonic instantons and instantonic
solitons (i.e.~instanton solutions trivially embedded in the
4+1~dimensional Yang-Mills theory) and profit from the vast knowledge
about zero-modes of the latter\footnote{It easy to check that
zero-modes of the instanton are zero-modes of the corresponding
instantonic soliton.}.  As we have already mentioned, there is a
one-to-one map between the moduli space of instantons and the moduli
space of dyonic instantons.  However, since the non-zero Higgs
expectation value of a dyonic instanton can not be obtained as a
(zero) mode deformation of the instanton, one cannot start from the
instanton moduli space. Instead one takes the ``instanton solution'' in
the Yang-Mills-Higgsed theory as a starting point.

It is well known that due to Derrick's theorem~\cite{Derrick:1964ww},
turning on the scalar field in the four-dimensional \emph{Euclidean}
Yang-Mills-Higgs theory destabilises the instanton solution against
collapse: this theory does not admit finite action solutions with
non-trivial scalar. The absence of \emph{exact} solutions does not,
however, imply that one cannot make use of \emph{approximate}
solutions of the field equations. These solutions are obtained by
expanding in small expectation values of the Higgs field at
infinity. The equations of motion for the four dimensional Euclidean
theory are
\begin{align}
\label{e:YMH1}
D_{i}F^{ij} &= [\Phi, D^{j}\Phi]\, ,  \\[1ex]
\label{e:YMH2}
D^2\Phi &= 0\, .
\end{align}
To lowest order in the Higgs expectation value the right hand side of
the equation~\eqn{e:YMH1} can be set to zero and the solution is then
given by~\eqn{e:sol-single} with $A_0=0$. This approximate solution
trivially lifts to a solution of 4+1~dimensional Yang-Mills-Higgs. We
will refer to this lifted solution, perhaps slightly inadequate, as
the ``constrained instantonic soliton''\footnote{In the
four-dimensional theory, these approximate solutions were used by
\dcite{Affleck:1981mp} to systematically compute instanton corrections
in the Higgs phase, despite the fact that an exact solution is not
available. His method of fixing the instanton to a given size by hand
is known as the ``constrained instanton method'', and the lowest order
term is obtained in the way described above.}.

Our starting point is thus the moduli space of constrained instantonic
solitons. Different points in this set are field configurations of
\emph{different} potential energy which solve equation of motion only
up to \emph{quadratic} order in the Higgs expectation value.  Hence, unlike the
situation for \halfbps dyons, an effective action that describes the
time evolution from one point in moduli space to the other necessarily
involves deformations which lead to a change in the potential energy
i.e.~deformations which are \emph{not} zero-modes. We will refer to
generic deformations that satisfy Gauss' law as \emph{modes}, and to
the subset of these which keep the potential fixed as \emph{zero
modes}.

Time evolution in the \emph{full} field theory can be approximated by
geodesic motion on the set of constrained instantons, provided the
time evolution in the full field theory is such that a system which is
initially tangent to this set stays near the set. In order for this to
be true, three requirements have to be met: the valley should be
completely known, the bottom of the valley has to be almost flat and
the edges surrounding the valley should be steep. The first condition
is necessary because there is in general no mechanism that restricts
the motion to a subset of the valley (unless, of course, there are
conservation laws at work). The second condition, together with the
third, ensure that the static force is small, so that the system never
develops enough kinetic energy to start climbing the mountains
that surround the valley. Note that the whole construction is completely
analogous to the one for \halfbps dyons, with the exception that the
valley is in that case precisely flat. In the general case, a slightly
tilted valley is allowed. The fact that the set of constrained
instantons is a complete set follows from the fact the scalar
equation~\eqn{e:YMH2} admits only a single solution given a self-dual
(i.e.~potential minimising) gauge-field background.

So let us now analyse the modes of the constrained instanton.  A
subset of the zero-modes of the instantonic soliton will also be
zero-modes of the constrained instantonic soliton, while other
zero-modes will be lifted by the presence of the non-zero
Higgs. However, all of these modes have to be taken into account since
the deformation by both types leaves us on the set of constrained
instantons.  Instanton zero-modes are very well studied in the
literature (for a review see for instance \dcite{Belitsky:2000ws}).
For the charge one SU(2) instanton, there are eight of them: four
corresponding to a change in position, one associated to the change of
the instanton size and three more having to do with the global
orientation of the instanton in the gauge group.  The associated
moduli are $y^i$, $\rho$ and the three angles $\theta^{a}$ ($a=1,2,3$)
which we have not considered so far. Our time-dependent ansatz is
therefore taken to be of the form
\begin{equation}
\label{e:sol-moving}
\begin{aligned}
A_0     &= \Big(\frac{\dot\theta^{a}}{H}\Big)\,T^{a}+\dot{y}^i \eta_{ij}^{a}\,\partial_j(\ln H)\, T^a\, ,  \\[1ex]
A_i     &= \eta_{ij}^{a}\,\partial_j(\ln H)\, T^a \, , \\[1ex]
\Phi    &= -\frac{v}{H}\, T^3 \, ,
\end{aligned}
\end{equation}
where the parameters $\rho$, $y^i$ and $\theta$ now depend on time.
The second term in $A_0$ can be deduced by using a Lorentz boost of
the static solution, or alternatively one can see that it corresponds
to the instanton translation zero-mode. The $\theta$-dependent terms
are  the global gauge embedding zero-modes, gauge transformed so that they
sit in $A_0$. When $\rho\, v$ and $\dot\rho$ are both small such that
their square can be neglected, one now indeed verifies
that~\eqn{e:sol-moving} satisfies Gauss' law: $D_i\partial_0 A_i=0$ by
itself and $D_iD^i A_0=0$ by virtue of the static equation of
motion. In general, we need $(\dot\rho)^2\ll 1$,
$(\rho\dot\theta)^2\ll 1$ and $(\rho v)^2 \ll 1$ for the approximation
to be valid (more about this issue in section~\ref{s:validity}).

As the translational zero-modes are relatively uninteresting for a
single dyonic instanton, let us set $\dot y^i=0$ and study the dynamics
of the size and group embedding parameters. We insert the
ansatz~\eqn{e:sol-moving} into the action~\eqn{e:action} and 
integrate over the space-like slices using the basic integral
\begin{equation}
I_0=\int\!{\rm d}^4x\, \frac{\rho^2 r^2}{(r^2+\rho^2)^4} = \tfrac{1}{3}\pi^2\, ,
\end{equation}
in terms of which we have
\begin{equation}
I_1 \equiv \int\!{\rm d}^4x\, \Big(D_i(1/H)\Big)^2 = 12\rho^2\, I_0\quad\text{and}\quad
I_2\equiv \int\!{\rm d}^4x\, \Big(\partial_\rho \partial_i \ln H\Big)^2 =
16\, I_0\, .
\end{equation}
With $(F_{0i})^2 = 3\,I_2\,\dot\rho^2 + I_1\,\dot\theta^2$ and $(D_i\Phi)^2= v^2\,I_1$,
the resulting moduli space action is
\begin{equation}
\label{e:Smsa}
S_{\text{moduli}} = 2\pi^2\,\int\!{\rm d}t\,
\Big(4\,\dot\rho^2 + \rho^2(\dot\theta^{a})^{2}- v^2\rho^2\Big)\, ,
\end{equation}
so that $\theta^{a}\in[0,4\pi)$. The $A_0$ component always leads to a
kinetic energy contribution, and together with the time variation of
$A_i$ they build up the first two terms in the action. These coincide
with those of the pure instanton. The third term, which is the most
interesting one, arises because we are in the Higgs phase, and
originates from the potential energy of the scalar\footnote{The
kinetic energy $\int (D_0\Phi)^2$ of the scalar field deserves special
attention, as it diverges logarithmically. This is very similar to the
situation for sigma model lumps (see Ward~\cite{Ward:1985ij},
Ruback~\cite{Ruback:1988sg} and Leese~\cite{Leese:1990hd}). Given that
it comes with a coefficient $v^2 \dot\rho^2$, one might be tempted to
interpret the divergence as a constraint on the dynamics
of~$\rho$. However, this would be incorrect for several reasons. First
of all, the term comes in at higher order in the approximation, and
could receive additional contributions. More importantly, we know that
for vanishing electric charge, $\dot\rho=0$ \emph{cannot} be a
solution to the equations of motion of the full theory, as it is
forbidden by Derrick's theorem. This seems to suggest that in the
approximation we make, this term can be ignored. Although a full
resolution of this puzzle is definitely still lacking, even for sigma
model lumps, we have verified that the cut-off method of Piette and
Zakrzewski~\cite{piet2} indeed confirms the correctness of our
procedure.}.  Note that, in contrast to earlier work where
potentials arose, we have derived it here simply by inserting the
explicit Higgs field from~\eqn{e:sol-moving} in the field theory
action.

As our $\theta$~zero modes are only valid to lowest order in
$\theta^a$, the action above is only correct when the angles are
small. For arbitrary angles, the factor $(\dot\theta^a)^2$ has to be
generalised to the SU(2) group volume. From now on, we will for
simplicity restrict our attention to a single angle~$\theta$
corresponding to the U(1) subgroup selected by the Higgs value at
infinity. This is a consistent truncation because one can check that
it restricts to a geodesic submanifold. For this particular angle the
kinetic term equals $\rho^2\dot\theta^2$ even for finite~$\theta$.

The equations of motion of the model~\eqn{e:Smsa} then consist of a
conservation law for \emph{angular momentum},
\begin{equation}
\label{e:angmom}
\dot\theta\,\rho^2 = L\, ,
\end{equation}
where $L$ is an arbitrary constant, together with the evolution
equation for the radius $\rho$,
\begin{equation}
\label{e:rhoeq}
- 8\,\ddot\rho + 2\,\rho \big(\dot\theta^2 - v^2 \big)=0\, .
\end{equation}
We will solve these in the general case in the next section, but there
are two special solutions which deserve separate attention. In the
first case we look at geodesics with non-zero angular momentum.  By
choosing the value of~$\theta$ appropriately one can cancel the
potential term completely. In this case, \eqn{e:angmom} fixes the
radius to be constant under time evolution,
\begin{equation}
\label{sol1}
\dot\rho=0\,, \quad \dot\theta = \pm v\, .
\end{equation}
This is the force balance discussed in the introduction. The geodesic
rotation in the direction chosen by the Higgs corresponds to an exact
dyonic instanton solution (with the signs corresponding to the freedom
in choosing $s'$ in~\eqn{e:sol-single}).  The electric charge of
dyonic instanton arises just as with the Julia-Zee dyon: it is
associated to kinetic energy in a compact direction (and corresponds
to the conserved angular momentum).

On the other hand, when the angular momentum is zero, the radius of
the instanton is driven to zero by the presence of the potential, and
to the extent to which this approximation is valid, one regains the
shrinking behaviour of the instantonic soliton in the Higgs
phase\footnote{Even in the pure Yang-Mills theory the instanton will
start to shrink if it is given an initial $\dot\rho\not=0$, since the
equation of motion~\eqn{e:rhoeq} reduces to $\ddot\rho=0$. A numerical
analysis of the shrinking behaviour in pure Yang-Mills was given by
\ddcite{Linhart:1999qb}{linh2}, who showed that the moduli space
approximation result $\rho=\rho_0+\dot\rho_0\, t$ gets modifications
in the full theory (the parameter $\rho$ becomes a function of $r$ and
the time evolution is modified as well). A similar analysis could
presumably be done in the Higgs phase, where corrections to the
$4\,\ddot\rho = - v^2 \rho$ evolution of the moduli space approximation
are expected to be found.}.

\end{sectionunit}
\begin{sectionunit}
\title{Dynamics of a single dyonic instanton}
\maketitle
\label{s:singledynamics}

\begin{figure}[t]
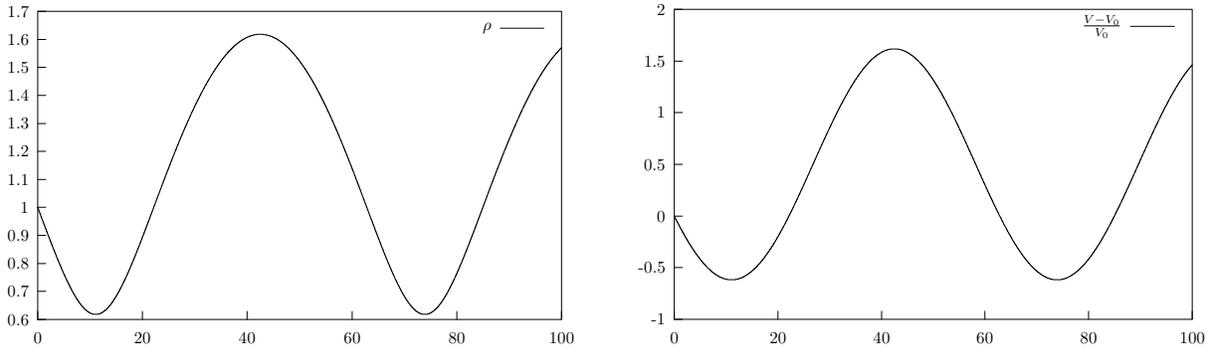

\begin{center}
\includegraphics*[width=.45\textwidth]{single_rho.eps}\quad\quad
\includegraphics*[width=.45\textwidth]{single_V.eps}
\end{center}
\caption{Typical oscillating orbit of a single dyonic instanton. The
radius is shown in the first plot, while the second one exhibits the
deviation of the potential from its initial value. Observe that the
potential is not constant even though the solution remains completely
within the regime of validity of the moduli space approximation. The
initial data are~\hbox{$\rho(t=0)=1$} and
\hbox{$\dot\rho(t=0)=-0.05$}, while the configuration is initially
rotating in the $T^3$~direction with critical
velocity,~\hbox{$\dot\theta=v$}.  Note that it is impossible to have a
critically rotating system when $\rho^2$ goes through its equilibrium
value; this follows directly by inserting $L=v\rho^2$ and $t=-t_0$
in~\eqn{e:rho_exact}.}
\label{f:singleorbit}
\end{figure}

The equations of motion~\eqn{e:rhoeq} and~\eqn{e:angmom} can be solved
exactly, and as we will see, the single dyonic instanton already shows
interesting dynamical behaviour, much richer than that encountered for
\halfbps solitons. Upon inserting the conservation law \mbox{$L=\dot\theta
\rho^2$} into the equation of motion for~$\rho$ we get
\begin{equation}
\label{e:rhodiffeq}
4\,\ddot \rho - \frac{L^2}{\rho^3} + v^2 \rho = 0\, .
\end{equation}
Multiplying by $\dot\rho$, this can be integrated once to
\begin{equation}
\label{e:cdef}
4\,(\dot\rho)^2 + \frac{L^2}{\rho^2} + v^2 \rho^2 = c^2\, .
\end{equation}
Upon further substitution of $\rho^2=x$ this becomes a standard
integral,
\begin{equation}
\int\!\frac{{\rm d}x}{\sqrt{-v^2 x^2+c^2x-L^2}} =
-\frac{1}{\sqrt{4L^2v^2-c^4}}
\arcsin\left(\frac{-2v^2 x + c^2}{\sqrt{4L^2v^2-c^4}}\right)\, .
\end{equation}
The solution is thus a harmonic oscillator for $\rho^2$. Eliminating
the constant $c$ in favour of the amplitude by using~\eqn{e:cdef}, the
expression for $\rho$ is
\begin{equation}
\label{e:rho_exact}
\rho = \sqrt{A\, \sin\big( v(t+t_0)\big) + \sqrt{\frac{L^2}{v^2}+A^2}}
= \sqrt{ 2A\, \sin^2\big( \tfrac{1}{2}v(t+\tilde t_0)\big) +
\sqrt{\frac{L^2}{v^2}+A^2} - A} \, .
\end{equation}
In the limit $L^2=0$ we thus recover the harmonic oscillator with
frequency $\tfrac{1}{2}v$, as expected from~\eqn{e:rhodiffeq}.
However, we should stress that as long as $\dot\theta\not=0$ (or in
other words, as long as $L^2\not=0$), the oscillating solution
described by~\eqn{e:rho_exact} does not go through zero. It is thus
fundamentally different from the pure constrained instantonic soliton:
whereas the solution with $\dot\theta=0$ goes through the singular
point $\rho=0$ in moduli space, and is therefore not necessarily a
valid orbit for arbitrary times, the oscillations found above are
genuine orbits that remain within the validity of the moduli space
approximation. A typical orbit is depicted in
figure~\ref{f:singleorbit} and compared with the \quarterbps dyonic
instanton in figure~\ref{f:threedplot}.

\begin{figure}[t]
\begin{center}
\includegraphics*[width=.5\textwidth, height=7cm]{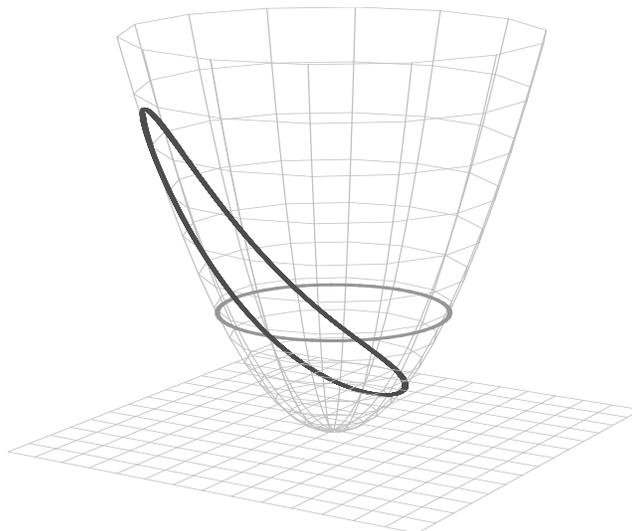}
\end{center}
\caption{Moduli space representation of the intrinsic dynamics of a
dyonic instanton. The $x$-$y$ plane is the moduli space of the
underlying instantonic soliton. The dyon is given by the circular orbit at
fixed height of the potential, with the angular velocity being related
to the electric charge (light curve). More kinetic energy can be added
to make the system oscillate around the \quarterbps configuration (dark
curve, corresponding to the data in figure~\ref{f:singleorbit}).}
\label{f:threedplot}
\end{figure}

It is also worth stressing that the generic solution does \emph{not}
stay at a fixed value of the potential. This is again different from
the pure dyonic instanton with $\dot\theta=v$, which does conserve the
potential and, because the potential gradient is exactly balanced by
the centrifugal force, has all other parameters independent of
time. The only quantity which is conserved in the general case is the
angular momentum $L$, related to the electric charge of the system.
\end{sectionunit}
\end{sectionunit}

\begin{sectionunit}
\title{Dynamics of two dyonic instantons}
\maketitle
\label{s:doubledynamics}
\begin{sectionunit}
\title{Metric and potential on moduli space}
\maketitle

With the understanding of the previous section at hand, we can now
start the main part of our program and construct the moduli space
approximation for the double dyonic instanton system. The modes
relevant for the computation of the metric are again the zero-modes of
the two-instanton solution. The resulting metric can be obtained from
the work of \dcite{Osborn:1981yf} (see also the work of
\dcite{Bruzzo:2000di} and \dcite{Bellisai:2000bc} who used the
hyperK\"ahler quotient construction). In addition, there will be a
potential term coming from the scalar sector. We will treat these two
parts of the calculation in turn.

Osborn's work~\cite{Osborn:1981yf} is focused on the calculation of
the two-instanton metric determinant, but his paper contains enough
intermediate results to extract the metric itself too. Details can be
found in appendix~\ref{e:metric2}. Let us first consider the case when
the two solitons are far apart. In this case the interpretation of the
quaternions used in the ADHM construction (see
appendix~\ref{e:metric2}) is straightforward: the magnitudes $|v_1|$
and $|v_2|$ are the sizes of the instantons, the unit normalised
quaternions $v_1/|v_1|$ and $v_2/|v_2|$ correspond to the SU(2)
angles, the centre of mass is $\rho$ and the positions are given by
$\rho+\tau$ and $\rho-\tau$ respectively.  First, consider the limit
$|\tau|\rightarrow\infty$. The metric then only receives contributions
from the first four terms of~\eqn{e:CpC1}. In vector notation, this
reads
\begin{equation}
{\rm d}s^2 = 4\Big( {\rm d}|v_1|\,{\rm d}|v_1| + |v_1|^2 (\Omega_1)^2 
+ {\rm d}|v_2|\,{\rm d}|v_2| + |v_2|^2(\Omega_2)^2
+ ({\rm d}\rho)^2 + ({\rm d}\tau)^2\Big)
\, .
\end{equation}
This correctly reduces to two copies of the flat metric of the single
dyonic instanton displayed in the first two terms of ~\eqn{e:Smsa}.

In case the group embedding angles are aligned the potential can be
obtained from the 't~Hooft ansatz for $\Phi$,
\begin{equation}
\Phi = -\frac{v}{H}\, ,\quad H=1+\frac{\rho_1^2}{(r-r_1)^2}
+ \frac{\rho_2^2}{(r-r_2)^2}\, .
\end{equation}
The resulting potential has been computed by~\dcite{Eyras:2000dg},
\begin{equation}
V = -(\rho_1^2+\rho_2^2) v^2\, .
\end{equation}
This is, as we will see, also the correct potential in the large
separation limit, and the equations decouple completely. However, the
full potential for the two-instanton case is more complicated and
depends on the relative gauge orientation.

When taking into account the $1/(\text{distance})^2$ corrections to
the metric, there are contributions from two terms in the appendix.
Interestingly, there are no corrections that depend on the impact
parameter (i.e.~that contain the angles of the quaternion $\tau$) at
this order. Adding the second term in~\eqn{e:dsds} and the first two terms
(squared) in~\eqn{e:smallk}, and also setting the centre of mass
coordinate to a constant, we get
\begin{multline}
{\rm d}s^2 = 4\,\Big( ({\rm d}v_1)^2 + ({\rm d}v_2)^2 + ({\rm
  d}|\tau|)^2 \Big)\\[1ex]
+ \frac{1}{|\tau|^2} \Big( ({\rm d}v_2 {\rm d}v_2)|v_1|^2
+({\rm d}v_1 {\rm d}v_1)|v_2|^2
+2\,(v_1 {\rm d}v_1)(v_2{\rm d}v_2)
-(v_1 {\rm d}v_2)^2 
-(v_2 {\rm d}v_1)^2
-2(v_1 v_2)({\rm d}v_1 {\rm d}v_2)
\Big)\\[1ex]
-\frac{1}{|\tau|^2}\big((v_1 {\rm d}v_2)-(v_2 {\rm d}v_1)\big)^2\, .
\end{multline}
The term in~\eqn{e:term1} which contains an epsilon term will only
contribute when the vectors $v_1$ and $v_2$ and their variations span
the four-dimensional space, i.e.~they contribute only to terms in the
metric which involve all three angles. A consistent truncation is
thus to set two of these angles identically zero, just as we did in
the single instanton case in section~\ref{s:singledynamics}. In this
case, it is convenient to use the parameterisation
\begin{equation}
v_1 = \rho_1\, ( \cos\theta_1 e_x + \sin\theta_1 e_y) \quad
\Rightarrow\quad 
{\rm d}v_1 = v_1(\theta_1)\, \frac{{\rm d}\rho_1}{\rho_1} +
v_1(\theta_1+\tfrac{\pi}{2})\,{\rm d}\theta_1\, ,
\end{equation}
and similarly for $v_2$. The metric can then be written as
\begin{multline}
\label{e:zeemetric}
{\rm d}s^2 = 4\,\Big( ({\rm d}\rho_1)^2 + ({\rm d}\rho_2)^2 +
\tfrac{1}{4}\big(\rho_1^2 + \rho_2^2\big) \big(({\rm d}\Theta)^2 + ({\rm d}\phi)^2\big)
+ \tfrac{1}{2}\big(\rho_1^2-\rho_2^2\big) {\rm d}\Theta{\rm d}\phi \Big)\\[1ex]
+ 4\,({\rm d}|\tau|)^2+ \frac{1}{|\tau|^2}\Big( {\rm d}(R\sin\phi) \Big)^2
- \frac{1}{|\tau|^2}\Big( 
\cos\phi (\rho_1 {\rm d}\rho_2 - \rho_2{\rm d}\rho_1)
+ R \sin\phi {\rm d}\Theta\Big)^2\, ,
\end{multline}
where we have introduced the variables
\begin{equation}
\begin{aligned}
\Theta &:= \tfrac{1}{2}(\theta_1+\theta_2)\, ,\\[1ex]
\phi   &:= \tfrac{1}{2}(\theta_1-\theta_2)\, ,\\[1ex]
R      &:= \rho_1\rho_2\, .
\end{aligned}
\end{equation}

For the potential, we essentially follow the work
of~\dcite{dore1}. This requires writing $(D_i\Phi)^2$ in terms of the
ADHM data and extracting the relevant parts in the large distance
approximation. The result is
\begin{equation}
\label{e:zeepotential}
V = v^2\Big(\big(\!\rho_1^2 + \rho_2^2\big) - \frac{1}{|\tau|^2}
R^2 \sin^2\phi\Big)\, .
\end{equation}
There are of course many terms at order $1/|\tau|^4$ which we have
ignored here.

The metric~\eqn{e:zeemetric} admits a Killing vector
$\partial/\partial\Theta$. As expected from the requirement that our
model can be supersymmetrised, one now indeed finds that the norm of
this Killing vector equals (up to an overall constant) the potential,
\begin{equation}
\left| \frac{\partial}{\partial\Theta}\right|^2 = \frac{V}{v^2}\, .
\end{equation}
Because the potential gradient contracted with the Killing vector
vanishes, i.e.
\begin{equation}
\frac{\partial V(x)}{\partial \Theta} = 0\, ,
\end{equation}
there is a conserved quantity associated to this Killing vector; it is
\begin{equation}
\begin{aligned}
L&=g_{\mu\Theta} \dot x^\mu \\[1ex]
&= \dot\Theta\Big( (\rho_1^2+\rho_2^2 -
\frac{1}{|\tau|^2}\rho_1^2\rho_2^2 \sin^2\phi\Big)
+ (\rho_1^2-\rho_2^2)\dot\phi - \frac{1}{|\tau|^2}\cos\phi\sin\phi
(\rho_1\dot\rho_2-\rho_2\dot\rho_1)\\[1ex]
&= \frac{\dot\Theta V}{v^2} + (\rho_1^2-\rho_2^2)\dot\phi - \frac{1}{|\tau|^2}\cos\phi\sin\phi
(\rho_1\dot\rho_2-\rho_2\dot\rho_1)\, .
\end{aligned}
\end{equation}
The effective action and the corresponding equations of motion can now
be written down easily using the metric~\eqn{e:zeemetric} and the
potential~\eqn{e:zeepotential}. We will refrain from spelling these
out, as they are rather lengthy, but some parts of them will be
displayed later when we discuss the actual dynamics.

\end{sectionunit}

\begin{sectionunit}
\title{Validity of the approximation}
\maketitle
\label{s:validity}
Before we analyse the orbits of the two-dyon system, let us make a few
comments on the validity of the approximation. We have already noted
that we are working in the limit in which $v$ and the velocities are
small, but since $v$ and $\dot\Theta$ are not dimensionless they have
to be compared to typical scales in the problem. The limit in which
our approximation holds is therefore
\begin{equation}
\begin{aligned}
\big(\rho(t)\,v\big)^2 &\ll 1\, , &&\quad\quad \big(\dot\rho(t)\big)^2 &\ll 1\, ,\\[1ex]
\big(\rho(t)\,\dot\Theta(t)\big)^2 &\ll 1\, ,&&\quad\quad \big(\dot\rho(t)\big)^2 &\ll 1\, , \\[1ex]
\big(\rho(t)\,\dot\phi(t)\big)^2 &\ll 1\, , &&\quad\quad (\rho^2/|\tau|^2)^2 &\ll 1\, .
\end{aligned}
\end{equation}
In these expressions, $\rho$ denotes any typical size parameter in the
problem.  The last condition could in principle be relaxed if one were
brave enough to include all the higher order $1/|\tau|$ corrections
into the metric and the potential calculation. The other conditions
are intrinsic to the moduli-space approximation and cannot easily be
avoided.
\end{sectionunit}

\begin{sectionunit}
\title{Analytical and numerical analysis of the orbits}
\maketitle
\label{s:nume}
We are now ready to analyse the solutions to the equations of motion
for the two-instanton system. From sections~\ref{s:modes}
and~\ref{s:singledynamics} we have learnt that there are two different
types of geodesics for the \emph{single} dyonic instanton: the BPS
dyonic instanton determined by
\begin{equation}
\label{geo1}
\rho_i = \text{constant}\, ,\quad  \dot{\theta}_{i}= v \, ,
\end{equation}
and the oscillating dyonic instanton, given by the more general
solution
\begin{equation}
\label{geo2}
\begin{aligned}
\rho^2_{i} &= A\sin\big(v(t+t_0)\big) + B\, ,\quad \dot{\theta}_{i} =
\frac{L}{\rho_i^2} \\[1ex]
B &\equiv \sqrt{ \frac{L^2}{v^2}+A^2}\, .
\end{aligned}
\end{equation}
In this section we will analyse the interaction and scattering
processes which as asymptotic states have: (i)~two BPS dyonic
instantons~\eqn{geo1} or (ii)~two oscillating dyonic
instantons~\eqn{geo2}.  As should be clear from the formulae for the
metric and potential, we are unable to give a completely analytic
treatment of the dynamics. The main qualitative features can however
be extracted with some help of numerical integration\footnote{The
computer programs used to produce these numerical solutions can be
obtained from \url{http://www.damtp.cam.ac.uk/user/kp229/di.html}.}.
\medskip

\noindent {\bf (i) two BPS dyonic instantons}:

The first check to perform is to verify that the double dyonic
instanton system at rest remains static at any finite separation. This
is required, since~\eqn{geo1} describes a BPS object, and hence it
satisfies the no-force condition. One indeed easily verifies that two
copies of solution~\eqn{geo1} for arbitrary values of the instanton
radii and arbitrary initial values of angle orientation $\theta_{i}$
$(i=1,2)$ solve the full equations of motion for the
metric~\eqn{e:zeemetric} with the potential~\eqn{e:zeepotential}.

When we kick the dyonic instantons towards each other, the dynamics
depends crucially on the value of the relative orientation~$\phi$ in
the gauge group. For $\phi=0$ or $\phi=\pi$, the sizes remain constant
and the group rotations remain undisturbed (this behaviour is in
contrast to that of ordinary \halfbps dyons, which when moving feel a
force due to the difference in the velocity dependence of the scalar
and vector forces). For the intermediate value $\phi=\pi/2$ the two
objects behave identically, with the radii slowly increasing as they
approach each other. In between these values, the behaviour of the
radii is more complicated.
\medskip

\noindent {\bf (ii) two oscillating instantons}:

Let us first consider two oscillating instantons which (when the
instantons are far apart) have identical $\rho$ oscillations. The
equation of motion for the separation $\tau$ following from
(\ref{e:zeemetric},\ref{e:zeepotential}) is given by
\begin{equation}
\label{e:eomtau}
\begin{aligned}
-8\,\ddot\tau - \frac{2v^2}{|\tau|^3} (\rho_1&\rho_2)^2
\sin^2\phi \\[1ex]
&- \frac{2}{|\tau|^3}\Big(\sin\phi
(\rho_1\dot\rho_2+\rho_2\dot\rho_1) + \rho_1\rho_2\,\dot\phi\,\cos\phi 
\Big)^2\\[1ex]
&+ \frac{2}{|\tau|^3}\Big( \cos\phi(\rho_1\dot\rho_2-\rho_2\dot\rho_1)
+ \rho_1\rho_2\,\dot\Theta\, \sin\phi\Big)^2=0\, .
\end{aligned}
\end{equation}
From this expression one notes that evolution with $\phi(t)=0$ and
identical radii, $\rho_1(t)=\rho_2(t)$, leads to a vanishing force and
hence $\tau=\text{constant}$ (inspection of the equation of motion for
$\phi$ shows that $\phi(t)=0$ can indeed be imposed for arbitrary
times). So we see that even though two identical and synchronised
oscillating dyonic instantons are not strict BPS states in the full
theory, they behave as BPS objects in our approximation. The
oscillating motion found in section~\ref{s:singledynamics} solves the
equations of motion of the effective action even when the solitons
are at finite separation.

More interesting dynamics occurs when one takes the relative angle for
the gauge embedding of the instantons to be nonzero
($\theta_1\not=\theta_2$, i.e.~$\phi\not=0$) while keeping the $\rho$
amplitudes the same. In this case one finds an attractive force
between the instantons for all initial values of~$\phi$.  A typical
orbit is depicted in figure~\ref{f:doublesync}. This behaviour is
difficult to extract from the equation of motion~\eqn{e:eomtau}
directly (due to the non-trivial oscillations of the radii). However,
we will see in the next section how an effective model can be derived
that predicts this behaviour.

Another, more complicated set of geodesics appears in case one has two
asymptotic dyonic instantons which oscillate with the same
$\rho$~amplitudes but with opposite phases. Depending on the initial
value of the angle~$\phi$ one finds attractive or, in contrast to the
situation before, repulsive behaviour. On top of this long time scale
motion, there are high-frequency oscillations in~$\tau$ which were not
present in the previous (equal phase) situation. The borderline case,
where only high-frequency oscillations remain, is hard to determine,
but we will see that again an effective model can come to the rescue.

\begin{figure}[t]
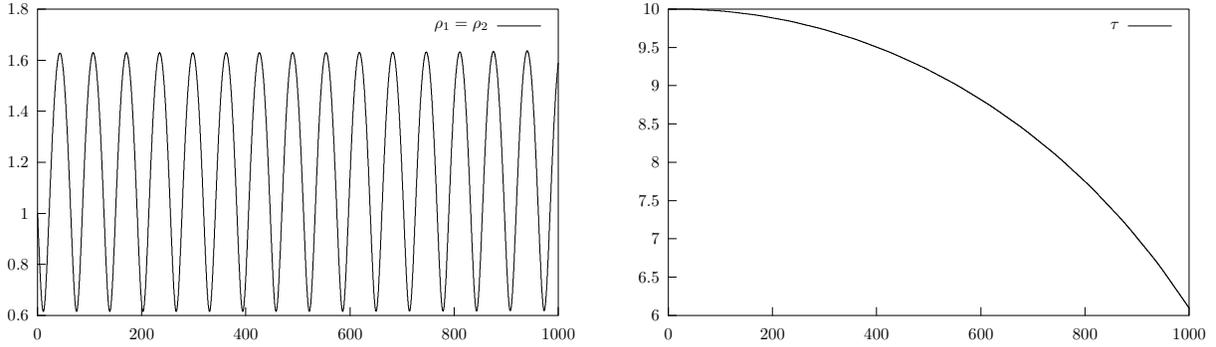

\begin{center}
\includegraphics[width=0.45\textwidth]{doublesync_rho.eps}\quad\quad
\includegraphics[width=0.45\textwidth]{doublesync_tau.eps}
\end{center}
\caption{Solution for which $\theta_1\not=\theta_2$, but the radii
 still satisfy $\rho_1=\rho_2$. This is only possible for $\phi=\pi/2$
 for all times, which implies that the group rotation of the two
 solitons is synchronised. The solitons feel an attractive force. The
 other initial values are $\rho_1=\rho_2=1$, $\tau=10$,
 $\dot\rho_1=\dot\rho_2=-0.05$. $A=1.118$}
\label{f:doublesync}
\end{figure}

Finally, when the condition of equal radii is dropped, the full
structure of the force between the solitons becomes visible.
Attractive, repulsive and oscillatory behaviour of $\tau$ are again
present, but the equations are generically too difficult to analyse
analytically. Typical orbits of this type, obtained using numerical
integration, are depicted in figure~\ref{f:doublerepel}
and~\ref{f:doubleattract}.

\begin{figure}[t]
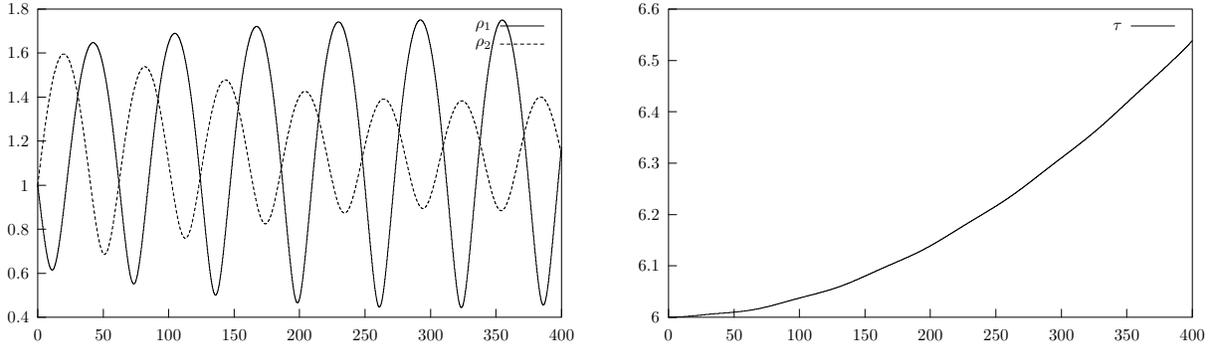

\begin{center}
\includegraphics[width=0.45\textwidth]{double_repel.0}\quad\quad
\includegraphics[width=0.45\textwidth]{double_repel.1}
\end{center}
\caption{Generic double soliton solution of the ``repulsive''
type. The initial data are $\tau=6$, $\rho_1=\rho_2=1$,
$\dot\rho_1=-\dot\rho_2=-0.05$ and $\dot\phi=0.7$. There are 
oscillations on top of the curve of $\tau(t)$, but they are too small
to be clearly visible in this plot.}
%
%
%
%
%
%
%
%
%
%
\label{f:doublerepel}
\end{figure}

\begin{figure}[t]
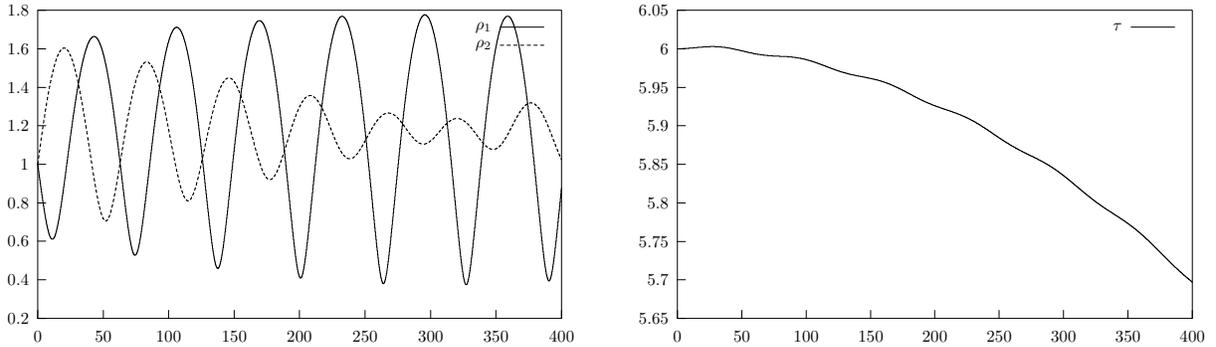

~\hfill\\[1ex]
\begin{center}
\includegraphics[width=0.45\textwidth]{double_attract.0}\quad\quad
\includegraphics[width=0.45\textwidth]{double_attract.1}
\end{center}
\caption{Generic double soliton solution of the ``attractive''
type. The initial data are similar to those of~\ref{f:doubleattract}
with the difference that now $\dot\phi=0.3$. Again, there are
oscillations on top of the $\tau$ curve which are very small.}
\label{f:doubleattract}
\end{figure}

\end{sectionunit}

\begin{sectionunit}
\title{Effective action for two oscillating dyonic instantons}
\maketitle 

Although the moduli space approach significantly simplifies the
analysis of soliton dynamics by getting rid of high-energy modes, one
can often still not do without numerical help when solving for the
low-energy behaviour. In the case of \halfbps dyons there is not much
else one can do, but for the \quarterbps solitons under consideration
here, we have seen that even in the low-energy regime there is
sometimes a separation of different scales.  For instance, the
attractive and repulsive orbits discussed in the previous section
exhibit high-frequency oscillation set by the scale of the instantons,
and low-frequency dynamics set by the scale of separation.  In the
present section, we will explain how one can simplify the analysis in
certain cases, by integrating out all but the lowest of the remaining
modes.

Let us first consider the case of two synchronised oscillating
instantons, $\rho_1(t)=\rho_2(t)$ with arbitrary $\phi$ and
$\theta$. As we have seen in the numerical analysis, this system
always exhibits attractive behaviour. Strictly speaking the amplitudes
of $\rho_1$ and $\rho_2$ are not equal in time, and not even constant
(the only case in which this is true is~$\phi=\pi/2$, given in
figure~\ref{f:doublesync}). However, since the rate of amplitude
decrease is much smaller than the rate of the decrease of the
instanton separation~$\tau$, one may approximate the instanton
oscillation during motion with the free one~\eqn{geo2}.  Numerical
analysis also shows that the relative phase $\phi$ oscillates with 
increasing amplitude. However as a first approximation we will take
$\phi$ to be a constant during motion.

So we proceed by inserting the $\rho_1(t)=\rho_2(t)$ given by
(\ref{geo2}) and $\theta=\text{const.}, \phi=\text{const.}$ into the
effective action derived from~\eqn{e:zeemetric}
and~\eqn{e:zeepotential}, and average over one period to arrive at
\begin{equation}
S_{\text{eff}} = \int\!{\rm d}t \left( 4\,\dot\tau^2 +
\frac{2\,v^2A^2\,\sin^{2}{\phi}}{\tau^2}\right)\, .
\end{equation}
As expected we see that potential is attractive for all values of the
angle $\phi$. The solution to the equation of
motion is easily found,
\begin{equation}
\tau = \sqrt{\tau_0^2 - \frac{v^2A^2 \sin^{2}(\phi)}{2\,\tau_0^2} t^2}\, .
\end{equation}
From this expression we can see that the Higgs expectation value~$v$
does not appear in the separation of the scales: the time scales for
$\rho$ and $\tau$ motion are $\omega_\rho=v$ and $\omega_\tau=v
A/\tau^2$ respectively. The agreement of the low-frequency mode
approximation with the full dynamics is excellent: when the analytical
result is compared with the numerical one of
figure~\ref{f:doublesync}, the errors are on the order of the width of
the curve.

It is important to stress that even though we have integrated out
high-frequency behaviour associated to the nonzero-modes, this is by
no means equivalent to ignoring them. While the latter corresponds to
keeping the radii fixed and leads to a vanishing force (as discussed
in part~(i) of the previous section) averaging over the nonzero-modes
leads to an attractive force.

Next we  consider two oscillating instantons with opposite phases
\begin{equation}
\label{e:rhos}
\begin{aligned}
\rho^2_{1} &= A\sin(vt) + B\, ,\quad&& \dot{\theta}_{1} &= \frac{L}{\rho_1^2} \\[1ex]
\rho^2_{2} &= -A\sin(vt) + B\, ,\quad&& \dot{\theta}_{2} &=
\frac{L}{\rho_2^2}\, . 
\end{aligned}
\end{equation}
By integrating the equation for
$\dot{\phi}=\frac{1}{2}(\dot{\theta_{1}}-\dot{\theta_{2}})$ one obtains
\begin{equation}
\label{e:phi}
\tan(\phi-\phi_{0}) = \frac{Av}{L}\cos(vt) \, .
\end{equation}
Inserting~\eqn{e:rhos} and~\eqn{e:phi} into the action constructed
from the metric~\eqn{e:zeemetric} and the
potential~\eqn{e:zeepotential}, one then finds
\begin{equation}
\label{effe}
\begin{aligned}
S &= \int 4(\dot{\tau})^2 + 
\frac{1}{\tau^2}\frac{L^2}{B^2-A^2\sin^{2}(vt)}
\frac{1}{1+\tan^2{\phi}}  \\[1ex] &\times 
\bigg{(} A^2 \sin^{2}(vt) \Big(1+
\frac{Av}{L}\cos(vt)\tan(\phi)\Big)^2 - B^2 \Big(\tan(\phi)-
\frac{Av}{L}\cos(vt)\Big)^2 \bigg{)} + \ldots 
\end{aligned}
\end{equation}
where dots corresponds to terms which after averaging produce constant
terms. By averaging the terms proportional to $1/\tau^2$ one obtains
again the effective force between two oscillating instantons. These
integrations are rather tedious to do analytically, so we have
restricted ourselves to a numerical integration for various values of
the $\phi_0$~parameter.  One finds that the effective
action~\eqn{effe} is indeed able to predict the attractive, repulsive
and oscillatory phases as obtained in the full numerical analysis of
section~\ref{s:nume}. Again, we should stress that integrating out the
nonzero-modes is crucially different from simply ignoring them, and
leads to much richer dynamics.

\end{sectionunit}
\end{sectionunit}

\begin{sectionunit}
\title{Discussion, conclusions and outlook}
\maketitle

In this paper we have exhibited several new features that arise when
the moduli space method is extended from \halfbps to \quarterbps
soliton dynamics. The main ingredient responsible for these new
features is the appearance of a potential, which we have explained
from the point of view of the underlying \emph{approximately static}
uncharged soliton. This approach has naturally led us to consider
dynamics for which the potential energy is not constant during
evolution. The nonzero-mode excitations can be taken into account in a
way that is consistent with the moduli space approximation, and
provide genuinely new types of behaviour. A second new feature is the
appearance of different scales even in the low-energy regime. This is
again intimately related to the presence of the potential: orbits
which involve motion up and down the potential ``superposed'' with
motion in the flat directions exhibit such a scale separation.  We
have shown how one can integrate out the high-frequency modes among
the excitations on moduli space, and arrive at a new effective action
for the remaining modes. We were able to illustrate these features on
the very explicit example of two dyonic instantons, as we managed to
write down an explicit expression for the metric and the potential of
this model.

For future research, several interesting questions remain
open. Firstly, the analysis of the dynamics is far from complete (we
have only considered head-on motion of the two solitons) and one may
want study more complicated orbits. Given the complexity of the
metric, we have also not been able to analyse full head-on scattering
that brings the solitons very close together. A particularly
interesting question is to understand the effect of the potential on
the $90^\circ$ scattering process familiar from dynamics of other
solitons.

On a more fundamental level, it would be interesting to study the
dynamical properties of the full supersymmetric extension of our
model, as well as the quantisation of it. Previous studies in this
direction (see for instance~\dcite{Bak:1999ip}) have focused on the
kinematical properties. Finally, there are other systems for which the
dynamics under influence of the potential can be analysed
explicitly. While most of the SU(3) monopole systems have a very
complicated metric (see~\dcite{Houghton:1999cm} for details), the
$(1,1)$ system is amenable to direct analytical analysis. Work in this
direction is in progress and will be published shortly.
\end{sectionunit}

\section*{Acknowledgements}

We thank Pierre van Baal, Jerome Gauntlett, Stefano Kovacs, Adam Ritz,
Paul Townsend and especially Conor Houghton and Nick Manton for
discussions, Ian Hawke for help with the numerical analysis and Johann
Strauss for sharing intuition about spinning instantons.

\vfill\eject
\appendix
\begin{sectionunit}
\title{Appendix: technical details}
\maketitle
\begin{sectionunit}
\title{Two-instanton metric}
\maketitle
\label{e:metric2}

The crucial ingredients for the computation of the metric on the
two-instanton moduli space can be found in~\dcite{Osborn:1981yf}. We
will here summarise the relevant details in a coordinate system that
is more suitable for our purposes; the reader is referred
to~\cite{Osborn:1981yf} for a detailed explanation of the formalism.

The square root $\Delta$ of the ADHM projection matrix will be
parametrised as
\begin{equation}
\Delta = a + b x 
\end{equation}
with 
\begin{equation}
a = \begin{pmatrix}
v_1 & v_2 \\
\rho+\tau & \sigma \\
\sigma & \rho-\tau 
\end{pmatrix}
\, ,\quad
b = \begin{pmatrix}
0 & 0 \\
\mathbb{1} & 0 \\
0 & \mathbb{1} 
\end{pmatrix}\, .
\end{equation}
The normalisation parameter $\nu$ appearing in $b^\dagger b = \nu \mathbb{1}$ is
therefore equal to one. The projector $P_\infty$ becomes
\begin{equation}
\label{e:Pinfty}
P_{\infty} =\lim_{x\rightarrow \infty} ( 1-b\nu^{-1} b^\dagger )
= \begin{pmatrix}
1 & 0 & 0 \\
0 & 0 & 0 \\
0 & 0 & 0 
\end{pmatrix}\quad
\text{so that}\quad
(1+P_\infty) = \begin{pmatrix} 2 & 0 & 0 \\ 0 & 1 & 0 \\ 0 & 0 & 1
\end{pmatrix}\, .
\end{equation}

We are now ready to solve the non-linear ADHM constraint $a^\dagger a
= \mu \mathbb{1}$. It leads to
\begin{equation}
\bar v_1 v_2 - \bar v_2 v_1 = 2(\bar \sigma \tau - \bar \tau \sigma)\,,
\end{equation}
which is necessary to make the off-diagonal components of $\mu$ vanish
(they are manifestly not proportional to the unit matrix in SU(2)
space). This allows us to solve for $\sigma$ in terms of the other
moduli:
\begin{equation}
\label{e:sigmasol}
\sigma = \tau \frac{\bar v_2 v_1 - \bar v_1 v_2}{4 |\tau|^2} + \alpha
\tau =: 
\tau(\frac{\Lambda}{4|\tau|^2} + \alpha)\, .
\end{equation}
We will set $\alpha=0$ in order to be able to use the interpretation
of the moduli as given by \dcite{dore1}. In order to make contact with
the work of \dcite{Osborn:1981yf}, we have to find a real function
$\lambda$ such that $\sigma$ above satisfies
\begin{equation}
\bar v_2 v_1 - 2\bar \tau \sigma = 2\lambda |\tau|^2\, .
\end{equation}
One finds that
\begin{equation}
\lambda = \frac{\bar v_2 v_1 + \bar v_1 v_2}{2 |\tau|^2}\, .
\end{equation}
We will see in a moment that this is proportional to $\delta^a{}_b$ in
spinor space. It is also useful to have a compact expression for $\Lambda$,
\begin{equation}
\Lambda = 4\,v_2^m v_1^n \bar\sigma_{mn} = -\bar\Lambda\, .
\end{equation}

The metric for the zero-modes can be obtained from (4.14) of
\dcite{Osborn:1981yf},
\begin{equation}
\label{e:zeromodemetric}
\langle C' , C \rangle = \underbrace{\langle \delta'a, \delta
a\rangle}_{:=\langle C',C\rangle_1} - 
\underbrace{4 k' k N_A^{-1}}_{:=\langle C',C\rangle_2}\, ,
\end{equation}
where the inner product of the first term is given by
\begin{equation}
\langle \delta'a, \delta a\rangle = \tr\big( \delta'a^\dagger (1+P_\infty)
\delta a\big)\, .
\end{equation}
The first term in \eqn{e:zeromodemetric} leads to a diagonal
contribution to the metric, which is easy to see by computing
\begin{multline}
\delta' a^\dagger (1+P_\infty) \delta a = \\[1ex] \begin{pmatrix}
2\,\delta' \bar v_1 \delta v_1 + (\delta'\bar\rho +
\delta'\bar\tau)(\delta \rho + \delta \tau) + \delta'\bar\sigma
\delta\sigma & \cdots \\[1ex]
\cdots & 2\,\delta' \bar v_2 \delta v_2 + (\delta'\bar\rho -
\delta'\bar\tau)(\delta \rho - \delta \tau) + \delta'\bar\sigma
\delta\sigma
\end{pmatrix}\, .
\end{multline}
The trace then eliminates the $\delta\rho\delta\tau$ cross-term, and
the result for the first term in the metric is
\begin{equation}
\label{e:CpC1}
\langle C',C\rangle_1 = 2\,\tr_2\big( \delta'\bar v_1 \delta v_1 + \delta'\bar v_2
\delta v_2 + \delta' \bar\rho \delta\rho +
\delta\bar\tau\delta\tau + \delta'\bar\sigma\delta\sigma \big)\, .
\end{equation}
The final step consists of eliminating the $\sigma$ in the last
term. The variations of $\bar\sigma$ and $\sigma$ are
\begin{equation}
\label{e:deltasigma}
\begin{aligned}
\delta'\bar\sigma &= \delta'\bar\Lambda \bar\tau \frac{1}{4|\tau|^2}+
\bar\Lambda\delta'\bar\tau \frac{1}{4|\tau|^2} - 
\bar\Lambda\bar\tau\delta'(\tau\bar\tau)\frac{1}{4|\tau|^4}\,  ,\\[1ex]
\delta\sigma &= \delta\tau \Lambda\frac{1}{4|\tau|^2} + 
\tau\delta\Lambda \frac{1}{4|\tau|^2} -
\tau\Lambda\delta(\tau\bar\tau)\frac{1}{4|\tau|^4}\, .
\end{aligned}
\end{equation}
The product thus leads to nine different terms for which we have to
work out the conversion of quaternions to vectors.
\begin{equation}
\label{e:dsds}
\begin{aligned}
(\delta'\bar\sigma)(\delta\sigma) &= 
\delta'\bar\Lambda \bar \tau \delta\bar\tau\Lambda\frac{1}{16|\tau|^4}+
\delta'\bar\Lambda \delta\Lambda\frac{1}{16|\tau|^2}-
\delta'\bar\Lambda \Lambda \delta(\tau\bar\tau)
\frac{1}{16|\tau|^4}\\[1ex]
&\quad +\bar\Lambda\delta'\bar\tau\delta\tau\Lambda\frac{1}{16|\tau|^4}+
\bar\Lambda\delta'\bar\tau\tau\delta\Lambda\frac{1}{16|\tau|^4}-
\bar\Lambda\delta'\bar\tau\tau\Lambda\delta(\tau\bar\tau)\frac{1}{16|\tau|^6}\\[1ex]
&\quad -\bar\Lambda\bar\tau \delta'(\tau\bar\tau) \delta\tau\Lambda
\frac{1}{16|\tau|^6}
-\bar\Lambda\delta'(\tau\bar\tau)\delta\Lambda\frac{1}{16|\tau|^4}
+\bar\Lambda\Lambda\delta'(\tau\bar\tau)\delta(\tau\bar\tau)\frac{1}{16|\tau|^6}
\, .
\end{aligned}
\end{equation}
In order to limit the number of $\sigma$ matrices appearing in the 
trace, it is convenient to rewrite $\delta(\tau\bar\tau)$ right
from the beginning in vector notation as
\begin{equation}
\label{e:deltatautau}
\delta(\tau\bar\tau) = 2{\mathbb 1}_2 (\tau\cdot\delta\tau)\, .
\end{equation}
In the main text we ignore all $1/|\tau|^4$ terms, so there is only
one term remaining,
\begin{equation}
\label{e:term1}
\begin{aligned}
\tfrac{1}{8}\tr(\delta'&\bar\Lambda\delta\Lambda) = \\[1ex]
&\phantom{+} (\delta'v_2\cdot\delta v_2) |v_1|^2
+ (v_1\cdot\delta' v_1)(v_2\cdot\delta v_2) + (v_1\cdot\delta v_1)(v_2\cdot\delta' v_2) 
+ (\delta' v_1\cdot \delta v_1) |v_2|^2\\[1ex]
& - (\delta' v_2\cdot v_1)(\delta v_2\cdot v_1)
- (\delta' v_1\cdot v_2)(\delta v_1\cdot v_2) - (v_2\cdot v_1)(\delta'v_1\cdot \delta v_2) 
- (v_2\cdot v_1)(\delta v_1\cdot\delta'v_2)\\[1ex]
& - \big(v_2^m \delta'v_1^n \delta v_2^k v_1^l + v_2^k
\delta v_1^l \delta'v_2^m v_1^n\big) \epsilon_{mnkl}\, .
\end{aligned}
\end{equation}

Now we focus on the second part, $\langle C',C\rangle_2$. For this we
have to compute
\begin{equation}
a^\dagger \delta a - (a^\dagger \delta a)^T = K = 
{\mathbb 1}_2\otimes\begin{pmatrix} 0 & k \\ -k & 0 \end{pmatrix}\, .
\end{equation}
For the matrix element $K_{12}$ one finds
\begin{equation}
K_{12}
 = \bar v_1 \delta v_2 - \bar v_2 \delta v_1 + 2\bar\tau\delta\sigma
- 2\bar\sigma\delta\tau\, .
\end{equation}
Inserting the solution~\eqn{e:sigmasol} of the ADHM constraint and
 using~\eqn{e:deltatautau}, this becomes
\begin{equation}
K_{12} = \tfrac{1}{2}\big( \bar v_1\delta v_2 + \delta\bar v_2 v_1 - \bar
v_2\delta v_1 - \delta\bar v_1 v_2\big)
-\tfrac{1}{2}\Big( \frac{\bar\Lambda}{|\tau|^2}\bar\tau\delta\tau
+ \delta\bar\tau \tau\frac{\Lambda}{|\tau|^2}\Big)\, .
\end{equation}
As expected this is indeed proportional to the identity matrix in
$SU(2)$: the terms that only have $v_i$ are proportional to
$\bar\sigma_m\sigma_n + \bar\sigma_n\sigma_m$, while the other ones
are proportional to $\bar\sigma_{mn}\bar\sigma_{kl}$, both of which
(by~\eqn{e:mnsymm} and~\eqn{e:sigma2sigma2} respectively) are
proportional to the identity. Rewriting $k$ in vector notation, we
find
\begin{equation}
\label{e:smallk}
k = (v_1\cdot\delta v_2) - (v_2\cdot \delta v_1)
- \frac{1}{|\tau|^2}\Big(\,\epsilon_{mnkl} v_2^m v_1^n \tau^k \delta\tau^l
-(v_2\cdot\tau)(v_1\cdot\delta\tau)
- \,(v_1\cdot\tau)(v_2\cdot\delta\tau)\Big)\, .
\end{equation}
From this we now have to compute $k'k (N_A)^{-1}$. This requires the
matrix $N_A$, which is given in~(3.30) of Osborn's paper,
\begin{equation}
N_A = |v_1|^2 + |v_2|^2 + 4 (|\tau|^2 + |\sigma|^2)
= |v_1|^2 + |v_2|^2 + 4 |\tau|^2 + |\bar v_2 v_1 - \bar v_2 v_1|^2
\, .
\end{equation}
The result for $k'kN_A^{-1}$ can now easily be read off.
\end{sectionunit}

\begin{sectionunit}
\title{Conventions and sigma algebra}
\maketitle
This appendix collects some useful expressions for Euclidean sigma
matrices (indices $m,n,\ldots$ take values $0,1,2,3$ but the zeroth
direction has positive signature too). The basic definitions in terms
of the Pauli matrices $\tau^i$ are
\begin{subequations}
\begin{align}
\sigma^m &:= (\mathbb{1}, i\tau^i)\, ,\\[1ex]
\bar\sigma^m &:= (\mathbb{1}, -i \tau^i)\, ,\\[1ex]
\sigma^{mn} &:= \tfrac{1}{4}\big( \sigma^m\bar\sigma^n -
\sigma^n\bar\sigma^m\big)\, ,\\[1ex]
\bar\sigma^{mn} &:= \tfrac{1}{4}\big( \bar\sigma^m \sigma^n -
\bar\sigma^n \sigma^m\big)\, ,\\[1ex]
\sigma^{mnp} &:= \tfrac{1}{2}\big( \sigma^m \bar\sigma^n \sigma^p 
- \sigma^p \bar\sigma^n \sigma^m\big)\, ,\\[1ex]
\bar\sigma^{mnp} &:= \tfrac{1}{2}\big( \bar\sigma^m \sigma^n \bar\sigma^p
- \bar\sigma^p \sigma^n \bar\sigma^m\big)\, .
\end{align}
\end{subequations}
The sigma matrices satisfy the algebra
\begin{subequations}
\label{e:sigmaalg}
\begin{align}
\sigma_l\bar\sigma_m &= 2\,\sigma_{lm} + \delta_{lm} \, ,\\[1ex]
\bar\sigma_l\sigma_m &= 2\,\bar\sigma_{lm} + \delta_{lm} \, .
\end{align}
\end{subequations}
and one can derive analogues for products of more sigma
matrices.
\begin{subequations}
\label{e:highersigma}
\begin{align}
\sigma_n\bar\sigma_{kl} &= \tfrac{1}{2}\sigma^{nkl} +
\delta^{n[k}\sigma^{l]} = \tfrac{1}{2}\epsilon^{nklq}\sigma_q
+ \delta_{n[k}\sigma_{l]}\, ,\\[1ex]
\sigma_m\bar\sigma_{nkl} &= \epsilon_{mnkl} - 2\epsilon_{nklq}\sigma_{mq}\, ,\\[1ex]
\sigma_{mn}\sigma_{kl} &= 
-\tfrac{1}{4}\delta_{mk}\delta_{nl}
+\tfrac{1}{4}\delta_{ml}\delta_{nk}
+ \tfrac{1}{4}\epsilon^{mnkl}\, ,\\[1ex]
\bar\sigma_{mn}\bar\sigma_{kl} &= 
-\tfrac{1}{4}\delta_{mk}\delta_{nl}
+\tfrac{1}{4}\delta_{ml}\delta_{nk}
- \tfrac{1}{4}\epsilon^{mnkl}\, .
\end{align}
\end{subequations}
The epsilon symbols (which satisfy $\epsilon_{0123}=1$) arise because
of the following duality flip relations,
\begin{subequations}
\begin{align}
\sigma_{mn} &= -\tfrac{1}{2}\epsilon_{mnpq} \sigma_{pq}\, ,\\[1ex]
\bar\sigma_{mn} &= \tfrac{1}{2}\epsilon_{mnpq} \bar\sigma_{pq}\, ,\\[1ex]
\sigma^{mnp} &= \epsilon^{mnpq} \sigma_{q}\, ,\\[1ex]
\bar\sigma^{mnp} &= -\epsilon^{mnpq} \bar\sigma_{q}\, .
\end{align}
\end{subequations}
Finally, one can of course trace these expressions to arrive at
\begin{subequations}
\begin{align}
\label{e:sigma2sigma2}
\tr_2\big( \sigma^{mnp} \big) &= 2\epsilon^{mnp0}\, ,\\[1ex]
\tr_2\big( \sigma_{mn}\sigma_{kl}\big) &=
-\tfrac{1}{2}\delta_{mk}\delta_{nl}
+\tfrac{1}{2}\delta_{ml}\delta_{nk}
+ \tfrac{1}{2}\epsilon^{mnkl}\, ,\\[1ex]
\tr_2\big( \bar\sigma_{mn}\bar\sigma_{kl}\big) &=
-\tfrac{1}{2}\delta_{mk}\delta_{nl}
+\tfrac{1}{2}\delta_{ml}\delta_{nk}
- \tfrac{1}{2}\epsilon^{mnkl}\, ,\\[1ex]
\label{e:mnsymm}
\tr_2\big( \sigma_m \bar\sigma_n + \sigma_n\bar\sigma_m\big) &=
4\delta_{mn}\, . 
\end{align}
\end{subequations}
\end{sectionunit}
\end{sectionunit}

\bibliography{moduli}

\begingroup\raggedright\begin{thebibliography}{33}
\expandafter\ifx\csname natexlab\endcsname\relax\def\natexlab#1{#1}\fi

\bibitem[Manton(1982)]{Manton:1982mp}
N.~S. Manton, ``A remark on the scattering of {BPS} monopoles'', {\em Phys.\
  Lett.} {\bf B110} (1982) 54--56.

\bibitem[Atiyah and Hitchin(1985)]{Atiyah:1985dv}
M.~F. Atiyah and N.~J. Hitchin, ``Low-energy scattering of nonabelian
  monopoles'', {\em Phys.\ Lett.} {\bf A107} (1985) 21.

\bibitem[Manton and Samols(1988)]{Manton:1988bn}
N.~S. Manton and T.~M. Samols, ``Radiation from monopole scattering'', {\em
  Phys.\ Lett.} {\bf B215} (1988) 559.

\bibitem[Julia and Zee(1975)]{juli1}
B.~Julia and A.~Zee, ``Poles with both magnetic and electric charges in
  {non-Abelian} gauge theory'', {\em Phys.\ Rev.} {\bf D11} (1975) 2227--2232.

\bibitem[Bergman(1998)]{Bergman:1997yw}
O.~Bergman, ``Three-pronged strings and {$1/4$-BPS} states in {$N=4$}
  {super-Yang-Mills} theory'', {\em Nucl.\ Phys.} {\bf B525} (1998) 104,
  \href{http://xxx.lanl.gov/abs/hep-th/9712211}{{\tt hep-th/9712211}}.

\bibitem[Alvarez-Gaum\'e and Freedman(1983)]{Alvarez-Gaume:1983ab}
L.~Alvarez-Gaum\'e and D.~Z. Freedman, ``Potentials for the supersymmetric
  nonlinear sigma model'', {\em Commun.\ Math.\ Phys.} {\bf 91} (1983) 87.

\bibitem[Tong(1999)]{Tong:1999mg}
D.~Tong, ``A note on {$1/4$-BPS} states'', {\em Phys.\ Lett.} {\bf B460} (1999)
  295, \href{http://xxx.lanl.gov/abs/hep-th/9902005}{{\tt hep-th/9902005}}.

\bibitem[Bak et~al.(2000{\natexlab{a}})Bak, Lee, and Yi]{Bak:1999vd}
D.~Bak, K.~Lee, and P.~Yi, ``Complete supersymmetric quantum mechanics of
  magnetic monopoles in {$N=4$} {SYM} theory'', {\em Phys.\ Rev.} {\bf D62}
  (2000){\natexlab{a}} 025009,
  \href{http://xxx.lanl.gov/abs/hep-th/9912083}{{\tt hep-th/9912083}}.

\bibitem[Bak et~al.(2000{\natexlab{b}})Bak, Lee, Lee, and Yi]{Bak:1999da}
D.~Bak, C.~Lee, K.~Lee, and P.~Yi, ``Low energy dynamics for {$1/4$ BPS}
  dyons'', {\em Phys.\ Rev.} {\bf D61} (2000){\natexlab{b}} 025001,
  \href{http://xxx.lanl.gov/abs/hep-th/9906119}{{\tt hep-th/9906119}}.

\bibitem[Gauntlett et~al.(2000)Gauntlett, Kim, Park, and Yi]{Gauntlett:1999vc}
J.~P. Gauntlett, N.~Kim, J.~Park, and P.~Yi, ``Monopole dynamics and {BPS}
  dyons in {$N=2$} {super-Yang-Mills} theories'', {\em Phys.\ Rev.} {\bf D61}
  (2000) 125012, \href{http://xxx.lanl.gov/abs/hep-th/9912082}{{\tt
  hep-th/9912082}}.

\bibitem[Gauntlett et~al.(2001)Gauntlett, Kim, Lee, and Yi]{Gauntlett:2000ks}
J.~P. Gauntlett, C.~Kim, K.~Lee, and P.~Yi, ``General low energy dynamics of
  supersymmetric monopoles'', {\em Phys.\ Rev.} {\bf D63} (2001) 065020,
  \href{http://xxx.lanl.gov/abs/hep-th/0008031}{{\tt hep-th/0008031}}.

\bibitem[Manton(1988)]{Manton:1988ba}
N.~S. Manton, ``Unstable manifolds and soliton dynamics'', {\em Phys.\ Rev.\
  Lett.} {\bf 60} (1988) 1916.

\bibitem[Manton and Merabet(1996)]{Manton:1996ex}
N.~S. Manton and H.~Merabet, ``{$\phi^4$ Kinks} - gradient flow and dynamics'',
  \href{http://xxx.lanl.gov/abs/hep-th/9605038}{{\tt hep-th/9605038}}.

\bibitem[Osborn(1981)]{Osborn:1981yf}
H.~Osborn, ``Semiclassical functional integrals for selfdual gauge fields'',
  {\em Ann.\ Phys.} {\bf 135} (1981) 373.

\bibitem[Lambert and Tong(1999)]{Lambert:1999ua}
N.~D. Lambert and D.~Tong, ``Dyonic instantons in five-dimensional gauge
  theories'', {\em Phys.\ Lett.} {\bf B462} (1999) 89,
  \href{http://xxx.lanl.gov/abs/hep-th/9907014}{{\tt hep-th/9907014}}.

\bibitem[Eyras et~al.(2001)Eyras, Townsend, and Zamaklar]{Eyras:2000dg}
E.~Eyras, P.~K. Townsend, and M.~Zamaklar, ``The heterotic dyonic instanton'',
  {\em JHEP\,} {\bf 05} (2001) 046,
  \href{http://xxx.lanl.gov/abs/hep-th/0012016}{{\tt hep-th/0012016}}.

\bibitem[Zamaklar(2000)]{Zamaklar:2000tc}
M.~Zamaklar, ``Geometry of the nonabelian {DBI} dyonic instanton'', {\em Phys.\
  Lett.} {\bf B493} (2000) 411--420,
  \href{http://xxx.lanl.gov/abs/hep-th/0006090}{{\tt hep-th/0006090}}.

\bibitem[Dorey et~al.(1996)Dorey, Khoze, and Mattis]{dore1}
N.~Dorey, V.~V. Khoze, and M.~P. Mattis, ``Multi-instanton calculus in {$N=2$}
  supersymmetric gauge theory'', {\em Phys.\ Rev.} {\bf D54} (1996) 2921--2943,
  \href{http://xxx.lanl.gov/abs/hep-th/9603136}{{\tt hep-th/9603136}}.

\bibitem[Bak and Lee(1999)]{Bak:1999sv}
D.~Bak and K.~Lee, ``Comments on the moduli dynamics of {$1/4$-BPS} dyons'',
  {\em Phys.\ Lett.} {\bf B468} (1999) 76,
  \href{http://xxx.lanl.gov/abs/hep-th/9909035}{{\tt hep-th/9909035}}.

\bibitem[Manton(1985)]{Manton:1985hs}
N.~S. Manton, ``Monopole interactions at long range'', {\em Phys. Lett.} {\bf
  B154} (1985) 397.

\bibitem[Derrick(1964)]{Derrick:1964ww}
G.~H. Derrick, ``Comments on nonlinear wave equations as models for elementary
  particles'', {\em J.\ Math.\ Phys.} {\bf 5} (1964) 1252--1254.

\bibitem[Affleck(1981)]{Affleck:1981mp}
I.~Affleck, ``On constrained instantons'', {\em Nucl.\ Phys.} {\bf B191} (1981)
  429.

\bibitem[Belitsky et~al.(2000)Belitsky, Vandoren, and van
  Nieuwenhuizen]{Belitsky:2000ws}
A.~V. Belitsky, S.~Vandoren, and P.~van Nieuwenhuizen, ``{Yang-Mills and
  D-instantons}'', {\em Class.\ Quant.\ Grav.} {\bf 17} (2000) 3521,
  \href{http://xxx.lanl.gov/abs/hep-th/0004186}{{\tt hep-th/0004186}}.

\bibitem[Ward(1985)]{Ward:1985ij}
R.~S. Ward, ``Slowly moving lumps in the {$CP^1$} model in (2+1) dimensions'',
  {\em Phys.\ Lett.} {\bf B158} (1985) 424.

\bibitem[Ruback(1988)]{Ruback:1988sg}
P.~J. Ruback, ``Sigma model solitons and their moduli space metrics'', {\em
  Commun.\ Math.\ Phys.} {\bf 116} (1988) 645.

\bibitem[Leese(1990)]{Leese:1990hd}
R.~Leese, ``Low-energy scattering of solitons in the {$CP^1$} model'', {\em
  Nucl.\ Phys.} {\bf B344} (1990) 33--72.

\bibitem[Piette and Zakrzewski(1996)]{piet2}
B.~Piette and W.~J. Zakrzewski, ``Shrinking of solitons in the
  (2+1)-dimensional sigma model'', {\em Nonlinearity} {\bf 9} (1996) 897--910.

\bibitem[Linhart(1999)]{Linhart:1999qb}
J.~M. Linhart, ``Fast blow up in the {(4+1)-dimensional} {Yang-Mills} model and
  the {(2+1)-dimensional} {$S(2)$} sigma model'',
  \href{http://xxx.lanl.gov/abs/math-ph/9909015}{{\tt math-ph/9909015}}.

\bibitem[Linhart and Sadun(????)]{linh2}
J.~M. Linhart and L.~A. Sadun, ``Fast and slow blowup in the {$S^2$} sigma
  model and $(4+1)$-dimensional {Yang-Mills} model'',
  \href{http://xxx.lanl.gov/abs/math-ph/0105024}{{\tt math-ph/0105024}}.

\bibitem[Bruzzo et~al.(2000)Bruzzo, Fucito, Tanzini, and
  Travaglini]{Bruzzo:2000di}
U.~Bruzzo, F.~Fucito, A.~Tanzini, and G.~Travaglini, ``On the multi-instanton
  measure for super {Yang-Mills} theories'',
  \href{http://xxx.lanl.gov/abs/hep-th/0008225}{{\tt hep-th/0008225}}.

\bibitem[Bellisai et~al.(2000)Bellisai, Fucito, Tanzini, and
  Travaglini]{Bellisai:2000bc}
D.~Bellisai, F.~Fucito, A.~Tanzini, and G.~Travaglini, ``Instanton calculus,
  topological field theories and {$N=2$} super {Yang-Mills} theories'', {\em
  JHEP\,} {\bf 07} (2000) 017,
  \href{http://xxx.lanl.gov/abs/hep-th/0003272}{{\tt hep-th/0003272}}.

\bibitem[Bak et~al.(2000)Bak, Lee, and Yi]{Bak:1999ip}
D.~Bak, K.~Lee, and P.~Yi, ``Quantum {$1/4$-BPS} dyons'', {\em Phys.\ Rev.}
  {\bf D61} (2000) 045003, \href{http://xxx.lanl.gov/abs/hep-th/9907090}{{\tt
  hep-th/9907090}}.

\bibitem[Houghton and Lee(2000)]{Houghton:1999cm}
C.~J. Houghton and K.~Lee, ``Nahm data and the mass of {$1/4$-BPS} states'',
  {\em Phys.\ Rev.} {\bf D61} (2000) 106001,
  \href{http://xxx.lanl.gov/abs/hep-th/9909218}{{\tt hep-th/9909218}}.

\end{thebibliography}\endgroup
\end{document}